\documentclass[useAMS,usenatbib,usegraphicx]{mn2e}
\usepackage{txfonts}
\title[Orbital Solution to Be/X-ray Binary IGR J01054-7253]{The Orbital Solution and Spectral Classification of the High-Mass X-Ray Binary IGR J01054-7253 in the Small Magellanic Cloud}
\author[L. J. Townsend et al.]{L. J. Townsend$^{1}$\thanks{E-mail: ljt203@soton.ac.uk (LJT)}, M. J. Coe$^{1}$, R. H. D. Corbet$^{2}$, V. A. McBride$^{1}$, A. B. Hill$^{3}$, A. J. Bird$^{1}$ \newauthor M. P. E. Schurch$^{4}$, F. Haberl$^{5}$, R. Sturm$^{5}$, D. Pathak$^{6}$, B. van Soelen$^{7}$, E. S. Bartlett$^{1}$ \newauthor S. P. Drave$^{1}$, A. Udalski$^{8}$\\
$^{1}$School of Physics and Astronomy, University of Southampton, Highfield, Southampton, SO17 1BJ, United Kingdom\\
$^{2}$University of Maryland Baltimore County, X-ray Astrophysics Laboratory, Mail Code 662, NASA Goddard Space Flight Center, Greenbelt, MD 20771, USA\\
$^{3}$Universit\'e Joseph Fourier - Grenoble 1 / CNRS, laboratoire d'Astrophysique de Grenoble (LAOG) UMR 5571, BP 53, 38041 Grenoble Cedex 09, France\\
$^{4}$Astrophysics, Cosmology and Gravity Centre (ACGC), Department of Astronomy, University of Cape Town, Rondebosch, Private Bag X1, Rondebosch 7701,\\ South Africa\\
$^{5}$Max-Planck-Institut f\"ur extraterrestrische Physik, Giessenbachstra\ss{}e, 85748, Germany\\
$^{6}$Faulkes Telescopes Project, School of Physics and Astronomy, Cardiff University, The Parade, Cardiff, CF24 3AA, United Kingdom\\
$^{7}$Department of Physics, University of the Free State, Bloemfontein, 9300, South Africa\\
$^{8}$Warsaw University Observatory, Aleje Ujazdowskie 4, 00-478 Warsaw, Poland}

\begin{document}

\date{Accepted 2010 August 18.  Received 2010 August 3; in original form 2010 July 1}

\pagerange{\pageref{firstpage}--\pageref{lastpage}} \pubyear{2010}

\maketitle

\label{firstpage}

\begin{abstract}

\noindent We present X-ray and optical data on the Be/X-ray binary (BeXRB) pulsar IGR J01054-7253 = SXP11.5 in the Small Magellanic Cloud (SMC). \textit{Rossi X-ray Timing Explorer (RXTE)} observations of this source in a large X-ray outburst reveal an 11.483 $\pm$ 0.002s pulse period and show both the accretion driven spin-up of the neutron star and the motion of the neutron star around the companion through Doppler shifting of the spin period. Model fits to these data suggest an orbital period of 36.3 $\pm$ 0.4d and $\dot{P}$ of (4.7 $\pm$ 0.3) $\times$ $10^{-10} \mathrm {ss}^{-1}$. We present an orbital solution for this system, making it one of the best described BeXRB systems in the SMC. The observed pulse period, spin-up and X-ray luminosity of SXP11.5 in this outburst are found to agree with the predictions of neutron star accretion theory. Timing analysis of the long-term optical light curve reveals a periodicity of 36.70 $\pm$ 0.03d, in agreement with the orbital period found from the model fit to the X-ray data. Using blue-end spectroscopic observations we determine the spectral type of the counterpart to be O9.5-B0 IV-V. This luminosity class is supported by the observed V-band magnitude. Using optical and near-infrared photometry and spectroscopy, we study the circumstellar environment of the counterpart in the months after the X-ray outburst.

\end{abstract}


\section{Introduction}

Be/X-ray binaries (BeXRBs) are systems in which a compact object orbits a massive, early type star that at some stage has shown evidence of line emission in the Balmer series. These stars are main sequence stars with a luminosity class of III to V and as such are a different sub-class of high-mass X-ray binary (HMXB) to the supergiant systems. To date there are no BeXRBs that contain a confirmed black hole \citep{bel09} meaning the detection of pulsed X-ray emission (caused by the strongly magnetized neutron star; NS) is an extremely robust tool for identifying such systems. The NS accretes via interactions with an extended envelope of material in the equatorial plane of the Be star. These systems typically have wide, eccentric orbits meaning most X-ray outbursts are due to the NS passing through periastron where the density of accretable material is greatest. These outbursts are denoted Type I outbursts and are generally in the luminosity range $10^{36}-10^{37}$erg s$^{-1}$ and last from a few days to a few weeks, depending on the binary period of the system. The less common Type II outbursts are brighter, $\gtrsim$\,$10^{37}$erg s$^{-1}$, and have no correlation with orbital phase. The cause of these outbursts is most likely an enlarged circumstellar disk that has grown to encompass part or all of the NS orbit, meaning accretion occurs for prolonged periods; from several days to a few months (see \citealt{stell86}, for further details of these outburst types). During phases of accretion, the pulse period of the pulsar is often seen to decrease, suggesting that large transfer of angular momentum from the companion to the NS is taking place. This is known as \textit{spin-up}. This phenomenon can be seen on both long-term (several years) and short-term (several days) time scales (e.g. SXP46.6, SXP144; \citealt{gal08}). Long-term changes in spin period are a complex mixture of spin-up from epochs of X-ray activity and spin-down from epochs of X-ray quiescence. Short term changes are most apparant under larger accretion rates and hence, high rates of angular momentum transfer. Therefore, observing these systems during a Type II outburst provides the best data to study accretion driven spin-up.

Interactions between the NS and the companion star are clearly important in determining the X-ray activity of any particular system, and so studying the circumstellar disk around the main sequence star is becoming an ever more prominent field of research. Optical monitoring by the Optical Gravitational Lensing Experiment (OGLE) on the 1.3m Warsaw telescope at Las Campanas observatory, Chile, has provided several years of I-band photometry of Magellanic Cloud stars. The relatively cool temperatures of the circumstellar disk in Be stars means that this material contributes strongly to the continuum emission in the near infrared (NIR). Thus, variability in the OGLE light curves of BeXRBs is thought to indicate changes in the structure or size of the circumstellar material. \citet{town10} and \citet{kem08} present examples of how flares in the optical light curves of some BeXRB systems can be associated with enhanced X-ray activity. In many cases, the variability is shown to be periodic or quasi-periodic; the latter thought to be associated with the growth and decay of the Be star's circumstellar disk. Temporal analysis often shows the periodic variability is caused by the presence of a compact object. The orbital periods of several BeXRBs have been discovered, or found to agree with a previously derived X-ray period, in this way (e.g. \citealt{schurch10}; \citealt{schmit06}). Variations in the size of the disk can also be studied through H$\alpha$ spectroscopy. \citet{gg06} present numerical models of circumstellar disks and show that the strength of the H$\alpha$ equivalent width is intrinsically related to the radius of the circumstellar material. These models rely on several observable parameters including the temperature of the star and the inclination of the disk to the observer; parameters that are obtainable by knowing the spectral type of the star and from the careful analysis of X-ray data like that outlined above. 

The Small Magellanic Cloud (SMC) is home to nearly 60 BeXRB systems \citep{coe08,cor08} as well as several candidate HMXB systems \citep{sht05}. The system that is the subject of this paper, IGR J01054-7253, was discovered as a new X-ray source in the direction of the wing of the SMC in June 2009 \citep{boz09}. It was later confirmed as a BeXRB system (see section 2 for details). In this paper we present extensive X-ray and optical data of this XRB. Section 2 gives details of the X-ray observations made and presents X-ray data of the outburst. Section 3 describes the orbital model fitting to several weeks of \textit{RXTE} data and presents the orbital solution for the system. Section 4 presents optical and IR data of the companion star. In section 5 we use Science Verification (SV) data from the new broadband spectrograph X-shooter on the Very Large Telescope (VLT) to spectrally classify the optical counterpart. Section 6 is a discussion of our findings. We review the implications of the accretion properties and orbital solution in a broader context and discuss what might be happening to the circumstellar disk in this system. We end with our conclusions in section 7.

\section{X-Ray Data}

\subsection{\textit{INTEGRAL}}

\begin{table}
  \caption{\textit{INTEGRAL} observations of the SMC and 47 Tuc in which IGR J01054-7253 was detected}
  \label{tab:sources}
  \centering
\begin{tabular}{|c|c|c|c|}
  \hline
  Revolution & MJD$_{start}$ & Exposure Time (ks) & Count Rate ($\mathrm{cts s}^{-1}$)\\
  \hline

  812 & 54989.2 & 63.3 & 0.4927\\
  813 & 54992.2 & 53.1 & 1.1790\\
  814 & 54995.2 & 60.8 & 1.1149\\
  815 & 54998.2 & 58.8 & 0.9149\\
  816 & 55001.2 & 62.1 & 0.8718\\
  817 & 55004.2 & 63.1 & 1.31728\\
  818 & 55007.2 & 40.2 & 1.35568\\

  \hline

\end{tabular}
\end{table}

\begin{figure*}
 \includegraphics[width=125mm,angle=90]{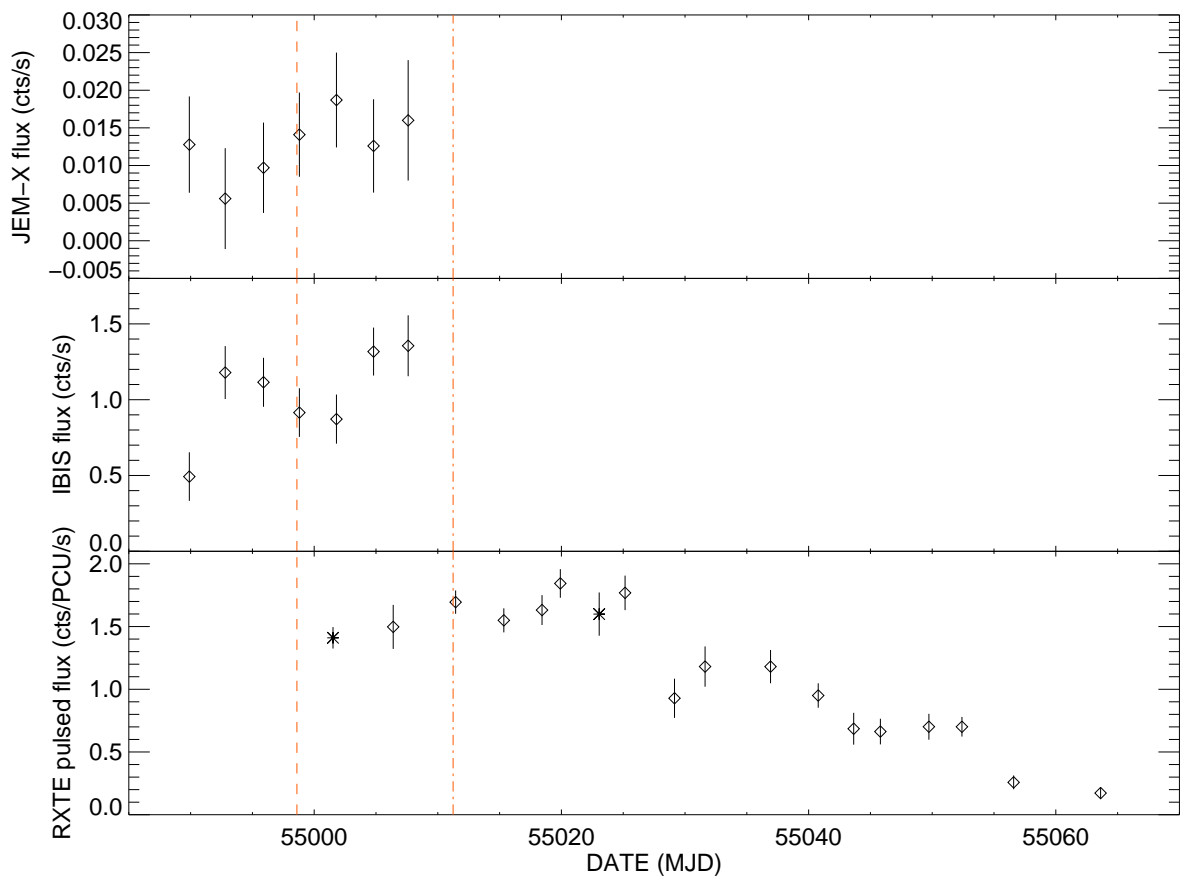}
  \caption{Combined \textit{INTEGRAL} and \textit{RXTE} light curve of IGR J01054-7253. The JEM-X flux (top panel) and \textit{RXTE} pulsed flux (bottom panel) are both in the 3--10 keV band, whilst the IBIS flux (middle panel) is in the 15--35 keV band. The two \textit{RXTE} points marked with a cross rather than a diamond are the observations used to extract spectra - see section 2.3 and Fig. \ref{fig:spect} for details. The vertical dashed and dot-dashed lines represent the time of \textit{Swift} and \textit{XMM-Newton} observations respectively.\label{fig:integral_lc}}
\end{figure*}

As part of a key programme monitoring campaign of the SMC and 47 Tuc, \textit{INTEGRAL} observed the SMC for approximately 90\,ks per satellite revolution ($\sim$3 days) between 2008 November 11 and 2009 June 25 (see \citealt{coe10} for more details). IGR J01054-7253 was detected by an \textit{INTEGRAL/IBIS} observation on MJD 54989 as a new X-ray source \citep{boz09}. Table \ref{tab:sources} gives the observation dates and exposures relating to the detection of this object. Unfortunately, these were the last seven observations of the key programme and Rev. 812 was the first observation of the SMC for nearly 45 days. As such we cannot be certain how long the source had been in outburst for before \textit{INTEGRAL} discovered it; it was in the \textit{RXTE} field of view, but only at a collimator response of approximately 0.15, making sources harder to detect. Likewise, we only detected the source with \textit{INTEGRAL} up to Rev. 818 which, as discussed below, was nearly two months before the source switched off.

Individual pointings (science windows) were processed using the \textit{INTEGRAL} Offline Science Analysis v.7.0 (OSA) \citep{gold03} and were mosaicked into revolution sky maps using the weighted mean of the flux in the 3--10\,keV (JEM-X) and 15--35\,keV (IBIS) energy ranges. Light curves in these energy bands were generated on science window ($\sim$2000\,s) and revolution time-scales. Fig. \ref{fig:integral_lc} shows the JEM-X (top panel) and IBIS (middle panel) revolution light curves of IGR J01054-7253. The JEM-X light curve appears to be mostly constant within errors, whereas some variation in the hard X-ray flux is apparant from the IBIS light curve. By eye it also appears that the hard and soft X-ray flux are anticorrelated with one another. However, plotting the hardness of the outburst in the ratio (15--35)/(3--10)\,keV against time showed no real evidence for a change in the hardness of the X-ray emission. Timing analysis on the \textit{INTEGRAL} light curves was not possible due to low count rates.

\begin{figure}
 \includegraphics[width=61mm,angle=90]{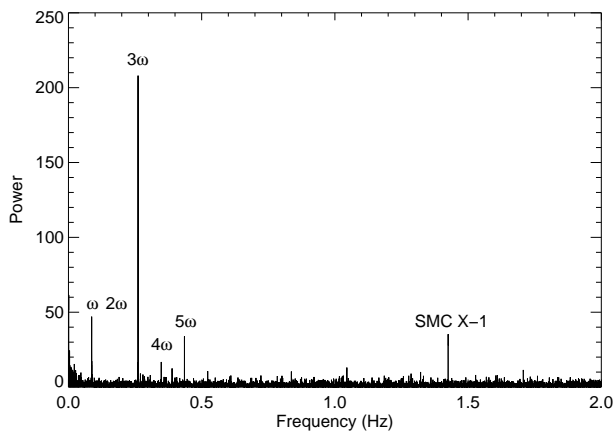}
  \caption{Power spectrum of IGR J01054-7253 showing the 11.5s spin period and the third, fourth and fifth harmonics. The second harmonic is not seen. SMC X-1 is also detected at higher frequencies.\label{fig:power}}
\end{figure}

\subsection{\textit{RXTE \& Swift}}

\begin{figure}
 \includegraphics[width=60mm,angle=90]{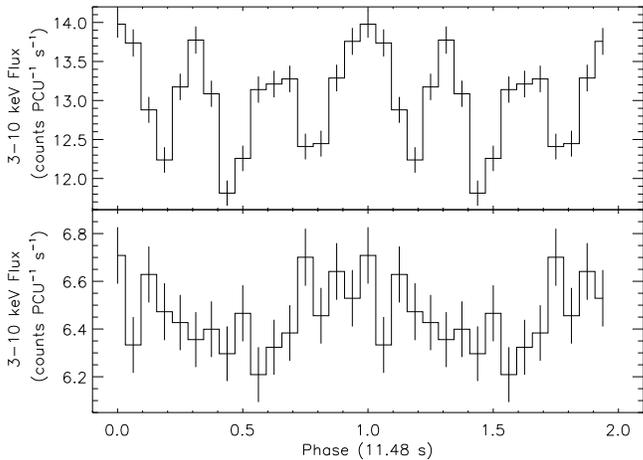}
  \caption{\textit{RXTE} 3--10 keV light curves folded at the observed spin period during the peak of the outburst (top panel: MJD 55019) and just before the end of the outburst (bottom panel: MJD55056). The pulse profiles have been smoothed and arbitrarily shifted in phase for viewing purposes.\label{fig:folded_lc}}
\end{figure}

\begin{figure}
\begin{center}
  \includegraphics[width=60mm,angle=90]{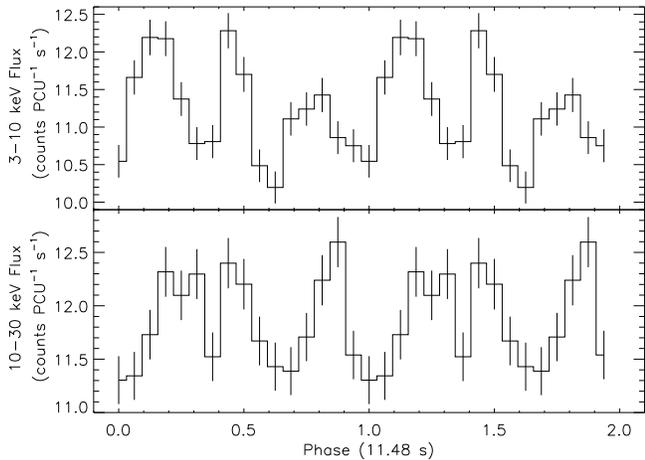}

\caption[]{\textit{RXTE} pulse profiles in the soft (3--10 keV) and hard (10--30 keV) energy bands from the observation on MJD 55023.}
\label{fig:rxte:pulseprofile}
\end{center}
\end{figure}

A \textit{Swift/XRT} follow up observation of the source took place on MJD 54998.6 aimed at refining the \textit{INTEGRAL} position and searching for pulsed X-ray emission. The observation revealed a precise position for this source of RA 01:04:41.41, Dec -72:54:04.6 (2000) with a 3.6 arcsec error circle \citep{coe09}. This position lies within 3.5 arcsec of the V=14.8 star [M2002] SMC 59977 which is now thought to be the optical counterpart. Further data on the counterpart are presented in sections 4 \& 5. Timing analysis performed on the XRT data is reported in \citet{coe09}. Those authors suggest that the most significant periodicity was 17.49s, although this is now known to be incorrect. The correct 11.48s period was not found in the XRT data until after \textit{RXTE} had found the correct period (\citealt{cor09}; IGR J01054-7253 will here-on be refered to as SXP11.5). A \textit{Swift} spectrum was also extracted and is discussed in \S 2.3. \textit{RXTE} observations began three days after the \textit{Swift} observation, once the source location had been refined, and lasted for just over two months. The PCA made a total of 18 exposures in Good Xenon mode during this time, meaning a time resolution of just under $1{\mu}\mathrm{s}$. The average exposure time was $\sim$6ks per observation. These data were then binned at 0.01s before being background subtracted to produce cleaned light curves in the 3--10 keV band. The cleaned light curves were then barycentre corrected and normalised to the number of PCU's on at each time interval during the observation. The \textit{RXTE} pulsed flux detections are presented in the lower panel of Fig. \ref{fig:integral_lc}. As can be seen, the pulsed flux peaked at around MJD55020 and steadily declined until dropping below the \textit{RXTE} sensitivity level (about $1 \times 10^{35} \mathrm{erg s}^{-1}$  at the SMC). The peak detected luminosity of the source was $1 \times 10^{37} \mathrm{erg s}^{-1}$ in the 3--10 keV band, using the average of several pulsed fraction measurements of 0.16. However, we stress that the pulsed fraction was only measured using a simple $\frac{max-min}{max+min}$ method and does not take into account the complex shape of the pulse profile; although the average value of 0.16 from the \textit{RXTE} observations is in agreement with the \textit{XMM-Newton} measurement (see \S 2.3) and the pulse shape does not change by much during the outburst, suggesting that these measurements are quite robust. Timing analysis was performed on these data using a Lomb-Scargle periodogram to search for the spin period of the NS. An example power spectrum that is representative of most of the observations is shown in Fig. \ref{fig:power}. The dominent periodicity detected was the third harmonic, with the fundamental period and the fourth and fifth harmonics detected at lesser powers. This behaviour is also seen in the phase-folded light curves. Fig. \ref{fig:folded_lc} shows two examples of the 3--10 keV light curves folded at the spin period observed by \textit{RXTE}. The profile in the upper panel shows how the third harmonic dominates the emission on MJD 55019 with an unusual triple structure; this shape is representative of all of the other profiles until the last few observations of the outburst. The profile in the lower panel shows how the emission geometry had changed into a profile dominated by the fundamental frequency by the end of the outburst, as shown by a much more single peak dominated profile. We also note here that a search over the entire 12 years of monitoring data from \textit{RXTE} reveals no previous detections of this source, although this may be due to a consistently low collimator response of less than 0.2. Another way of exploring the emission on the NS surface is to divide an observation up into several energy bands and fold each light curve on the detected period, giving energy dependent phase-folded light curves. Fig. \ref{fig:rxte:pulseprofile} shows the 3--10 keV (soft) and 10--30 keV (hard) \textit{RXTE} folded light curves from the observation made on MJD 55023. The soft profile is very similar to that shown in the top panel of Fig. \ref{fig:folded_lc}, demonstrating that there was little variation in the emission in this energy range. Comparing the soft and hard profiles however, shows evidence for a change in structure; the 3rd peak in the profile changes from being smaller than the other two to a comparible height. This is possible evidence that the emission mechanism is different at different energies. This will be discussed further in section 6.

\begin{figure}
\begin{center}
  \resizebox{0.95\hsize}{!}{\includegraphics[clip=,angle=0]{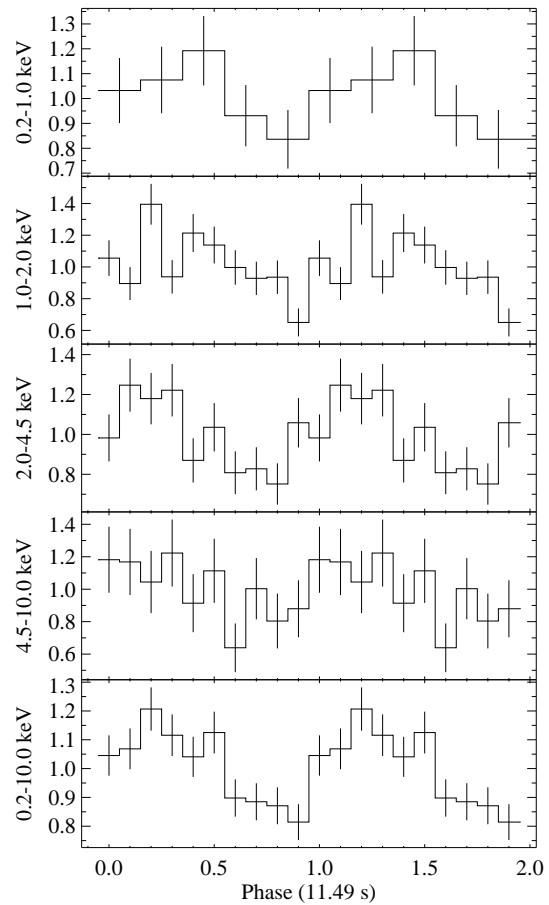}}

\caption[]{\textit{XMM-Newton} pulse profile of SXP11.5 derived from MOS 1 data in different energy bands. The profiles are background subtracted and then plotted as a ratio to the mean count rate. The mean count rates for the energy bands used are (from top to bottom): (1.24, 3.44, 2.88, 1.18 and 8.75) x 10$^{-2}$ cts/s.}
\label{fig:xmm:pulseprofile}
\end{center}
\end{figure}

\subsection{\textit{XMM-Newton}}

SXP11.5 was serendipituously detected during one observation of the \textit{XMM-Newton} \citep{jan01} large programme SMC survey \citep{habp08} on MJD 55011.24--55011.59 (ObsID 0601212601). The source was visible only in CCD7 of EPIC-MOS1 \citep{turn01} at the very rim of the detector, where a light path outside the nominal field of view (FoV) exists due to a gap in the instrument structure (cf. \citealt[][ Fig. 11]{turn01}). We note, that the calibration for this region might not be as advanced as for the rest of the detector.

We used \textit{XMM-Newton} SAS10.0.0\footnote{Science Analysis Software (SAS), http://xmm.esac.esa.int/sas/} for data processing. We identified sources in the FoV for astrometric boresight correction ($\Delta$RA=-3.49", $\Delta$Dec=0.63") and obtained for SXP11.5 a revised \textit{XMM-Newton} position of RA 01:04:41.60, Dec -72:54:04.5, which agrees very well with the position determined by \textit{Swift}. For the systematic error we estimate $\sim$4 arcsec, since the source lies at a large off-axis angle of $\sim$17 arcmin.

For data reduction we selected events with {\tt PATTERN$\le$12} in an elliptical extraction region, placed on the source, and a box, lying on the point source free part of the light-leak region, respectively. Filtering of periods with high background was not necessary, since soft proton flares were at a quiescent level, yielding a net deadtime corrected exposure of 24.3 ks. For spectral analysis we used only events with {\tt flag = 0xfffeffff} (similar to {\tt flag=0}, but including events outside the nominal FoV). The spectrum was binned to contain $\ge$20 counts bin$^{-1}$ and response matrices and ancillary files were created using the SAS tasks {\tt rmfgen} and {\tt arfgen}.

The power density spectrum of this observation exhibits a peak at the period found by \textit{RXTE} at 0.087 Hz. Using a Bayesian periodic signal detection method \citep{greg96} yields a pulse period of $(11.483\pm 0.003)$ s (1$\sigma$ error). The \textit{XMM-Newton} detection independently confirms both the position of this source and its spin period. The folded pulse profile in different energy bands is plotted in Fig. \ref{fig:xmm:pulseprofile}. The shape of the 0.2--10.0 keV profile is relatively sinusoidal and a calculated pulsed fraction of 0.16 agrees with \textit{RXTE} measurements. Although the third harmonic is not seen in the \textit{XMM-Newton} data, probably because it is close to the detector timing resolution of 2.6s, the profiles show that there is very little variation in the X-ray emission amongst these energy bands.

\begin{figure}
 \includegraphics[width=60mm,angle=270]{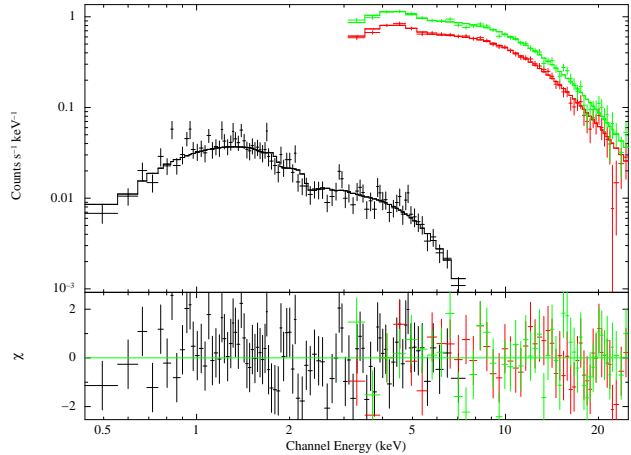}
  \caption{\textit{RXTE} \& \textit{XMM-Newton} spectra recorded during the outburst. The \textit{XMM-Newton} spectrum was taken on MJD 55011 and is shown in black. The \textit{RXTE} spectra were taken on MJD 55001 \& 55023 (red and green respectively). All three spectra were fit simultaneously with an absorbed power-law with high-energy cutoff allowing only a constant normalisation factor between the three spectra under the assumption that the spectral shape does not change. The model fit is presented in the text.\label{fig:spect}}
\end{figure}

Fig. \ref{fig:spect} presents the simultaneous fit to the \textit{XMM-Newton} (black) and \textit{RXTE} spectra (red and green taken 10 days before and 12 days after the \textit{XMM-Newton} detection respectively - see Fig. \ref{fig:integral_lc}). The two \textit{RXTE} spectra were chosen as they are the ones taken closest in time to the \textit{XMM-Newton} spectrum that contain no other active pulsar in the field of view, based on the \textit{RXTE} power spectrum for that observation. The spectra were modelled with an absorbed power-law with a high-energy cutoff allowing only a constant normalisation factor between the three spectra with the assumption of no change in the spectral shape (only intensity changes). The flux during the \textit{XMM-Newton} observation was apparently significantly lower (although there is some uncertainty due to the far-offaxis source position). We fixed the Galactic photoelectric absorption at an N$_{\rm H}$ = 6$\times 10^{20}$cm$^{-2}$ \citep{dic90}, whereas the SMC column density with abundances at 0.2 for metals was a free fit parameter. We obtained a best-fit with $\chi^2/{\rm dof} = 184/179$ with the best-fit parameters: N$_{\rm H, SMC} = 2.30^{+0.63}_{-0.60}\times 10^{21}$cm$^{-2}$, $\Gamma = 1.06^{+0.05}_{-0.05}$, cutoff energy = $9.24^{+1.87}_{-0.98}$ keV, folding energy = $22.0^{+4.2}_{-4.2}$ keV and
a detected flux of $(7.81\pm 0.57)\times 10^{-12}$ ergs cm$^{-2}$s$^{-1}$ in the 3--10.0 keV band. Assuming a distance of 60 kpc, this corresponds to an unabsorbed \textit{XMM-Newton} luminosity of $L_{3-10.0}=(3.35\pm 0.25)\times 10^{36}$ ergs s$^{-1}$. The luminosity of the two \textit{RXTE} observations found using the normalisation fit are $L_{3-10.0}=(1.79\pm 0.07)\times 10^{37}$ ergs s$^{-1}$ and $L_{3-10.0}=(2.54\pm 0.11)\times 10^{37}$ ergs s$^{-1}$ for the red and green spectra respectively. These agree nicely with the value obtained using the pulsed flux light curve and the pulsed fraction. The \textit{XMM-Newton} value is much lower than the values found in the \textit{RXTE} observations. It is unclear how much of this difference is due to source variability and how much is due to the unusually far off-axis position of the source on the MOS detector. For comparison, this model was also fit to the spectrum from the \textit{Swift} observation. The luminosity was found to be $L_{3-10.0}=(9.51\pm 0.48)\times 10^{36}$ ergs s$^{-1}$. This value is a factor of 2 smaller than the \textit{RXTE} value measured just 3 days later, but could be explained by variability within the source at the start of the outburst.

SXP11.5 was also in the FoV of a previous \textit{XMM-Newton} observation on MJD 54010.99--54011.26 (ObsID 0402000101), when no significant X-ray emission at this position was detected. Analysing the EPIC-pn data, we found a 3$\sigma$ upper limit of 0.002 cts s$^{-1}$ for the 0.2--12.0 keV energy band, which, assuming the same spectrum as above, corresponds to 
an unabsorbed luminosity limit of $L_{0.2-10.0}\le 6.4\times 10^{33}$ ergs s$^{-1}$.

\section{Orbital Solution to IGR J01054-7253}

\begin{figure}
 \includegraphics[width=64mm,angle=90]{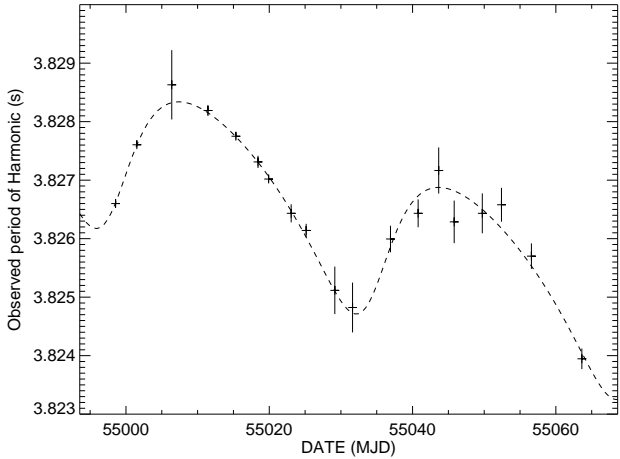}
  \caption{Period of the third harmonic of SXP11.5 as detected by \textit{RXTE} during the two months of pointed observations. The final two data points are one third of the detected fundamental period as the third harmonic had disappeared by this point and the first data point is one third of the period detected by \textit{Swift} (see text). Overplotted is the model fit to the data as described in the text.\label{fig:orbit}}
\end{figure}

It became apparent after the first few \textit{RXTE} observations that there were both spin-up and spin-down trends in the pulse period of SXP11.5, indicating that we were seeing the motion of the NS around the companion star through Doppler shifting of the spin period, similar to that seen in SXP18.3 \citep{schurch09} and GRO J1750-27 \citep{shaw09}. Fig. \ref{fig:orbit} shows the detections of the third harmonic of the pulse period during the outburst and the associated errors. Overplotted is the model fit to the data as discussed later. The third harmonic was used to fit the model because it was several times more powerful in the power spectrum than the fundamental frequency (Fig. \ref{fig:power}) and so the associated errors are smaller. The error bars increase in size towards the end of the outburst as the pulsed emission became much weaker as the pulsar switched off. The final two data points are one third of the detected fundamental period as the third harmonic had disappeared by this point (see discussion) and the first data point is the \textit{Swift} detection, which is also one third of the detected period as, like the MOS-1 detector on \textit{XMM-Newton}, we were unable to detect the third harmonic due to longer binning needed with such a low count rate source. Due to the length of the gaps between observations, it was not possible to match phase between them and do more accurate pulse arrival time analysis.

As mentioned, the oscillatory nature of the spin period is most easily explained by the Doppler shifting of the X-ray light from the NS by the binary orbital motion in the system. However, a general spin-up trend, most likely caused by angular momentum transfer during the accretion of matter onto the NS, is also apparent in the data. Consequently the observed data were fit with a simple spin-up model given by

\begin{equation}
 P(t) = P(t_{0}) + \dot{P}(t - t_{0}) - \frac{\ddot{P}(t - t_{0})^{2}}{2}
  \label{equ:spin}
\end{equation}

\noindent where \textit{P} is the spin period of the neutron star, $\dot{P}$ and $\ddot{P}$ are the spin-up and change in spin-up, and t$_{0}$ and t are the start time and time since the start time over which the model is fit. This model was convolved with a standard orbital model which calculates the line of sight velocity of the NS. This combined model iterates over some input parameters until a best fit for a Levenberg-Marquardt least-squares fit is achieved. Standard values for other HMXB systems were used for initial parameters that were unknown. After first running a simplified version of this model that assumed a circular orbit, it was clear that an acceptable fit could only be achieved by including an eccentricity component and excluding $\ddot{P}$ as the data were not sufficient to fit the second derivative. The final fit gave an orbital period of 36.3 $\pm$ 0.4d and a $\dot{P}$ of (4.7 $\pm$ 0.3) $\times 10^{-10} \mathrm{ss}^{-1}$ \citep{town09}. The model fit to the data is shown in Fig. \ref{fig:orbit} and the various parameter values are displayed in Table \ref{tab:orbit} along with the reduced chi-squared value.

\begin{table}
  \caption{The orbital parameters for IGR J01054-7253 from the analysis of 3--10 keV \textit{RXTE} PCA data.}
  \label{tab:orbit}
\begin{tabular}{|l|c|l|}
  \hline
  Parameter &  & Orbital Solution \\
  \hline
  Orbital period & $\textit{P}_{orbital}$ (d) & $36.3\pm0.4$ \\
  Projected semimajor axis & $\textit{a}_{x}$sin{\it i} (light-s) & $167\pm7$ \\
  Longitude of periastron & $\omega$ ($^{o}$) & $224\pm10$ \\
  Eccentricity & $\textit{e}$ & $0.28\pm0.03$ \\
  Orbital epoch & $\tau_{periastron}$ (MJD) & $ 55034.3\pm1.0$ \\
  Spin period & $\textit{P}$ (s) & $11.48143\pm0.00001$ \\
  First derivative of $\textit{P}$ & $\dot{P}$ ($\mathrm{ss}^{-1}$) & $(-4.67\pm0.31)\time10^{-10}$ \\
  Goodness of fit & $\chi^{2}_{\nu}$ & 0.82 \\
  \hline

\end{tabular}
\end{table}

\section{Optical and IR Data}

In this section, we present spectroscopic and photometric data in the optical and NIR wavebands from a variety of ground-based telescopes. The aims of this were to spectrally classify the optical counterpart and to explore the changes in the circumstellar disk leading up to, during and after the X-ray outburst.

\subsection{Observatories and Instrumentation}

\begin{itemize}

\item OGLE III - The OGLE project has been steadily monitoring millions of stars in the Magellanic Clouds for the past 12 years (see \citet{uda97} and \citet{syzm05} for more details on the OGLE instrumentation and catalogue). Photometric images in the I-band are taken almost every night whilst the SMC is visible; these images now make up a decade long database that includes most of the HMXB systems in the SMC.

\item Faulkes Telescope - FT South, located at Siding Spring, Australia is a 2m, fully autonomous, robotic Cassegrain-type reflector which employs a Robotic Control System (RCS) \citep{tsapras09}. The telescope was used both in “Real Time Interface” (RTI) mode and “Offline” mode for the observations of SXP11.5. All the observations were pipeline-processed which does the flat-fielding and de-biasing of the images.

\item 1.9m Radcliffe telescope at the South African Astronomical Observatory (SAAO) - Photometry was done using the SAAO CCD at the Cassegrain focus. The spectra were obtained using the unit spectrograph combined with a 1200 l/mm grating and the SITe detector at the Cassegrain focus. Data reduction was performed using standard IRAF packages.

\item 1.4m Infrared Survey Facility (IRSF) telescope at SAAO - The IRSF is a Japanese built telescope designed specifically to take simultaneous photometric data in the J, H \& $K_{s}$ bands with the SIRIUS (Simultaneous three-colour InfraRed Imager for Unbiased Survey) camera \citep{nag99}. Data reduction was performed using the dedicated SIRIUS pipeline based on the National Optical Astronomy Observatory's (NOAO) IRAF software package. The pipeline was provided by Yasushi Nakajima at Nagoya University, Japan. This performs the necessary dark subtraction, flat fielding, sky subtraction and recombines the dithered images. Photometry on the reduced images was performed using standard IRAF routines.

\item X-shooter - The first of the second generation VLT instruments, X-shooter \citep{dod06}, is a three arm, single object echelle spectrograph for the Cassegrain focus of one of the VLT UT's. The instrument simultaneously covers the wavelength range 300-2400 nm at resolving powers R = $\frac{\lambda}{\Delta\lambda}$ = 5100, 8800 and 5600 in the UVB ($\Delta\lambda$ = 300-550 nm), VIS ($\Delta\lambda$ = 550-1015 nm), and NIR arms ($\Delta\lambda$ = 1025-2400 nm) respectively. However, the resolution obtained is dependent on the slit width and seeing conditions.

\end{itemize}

\subsection{Photometry}

\begin{figure}
 \includegraphics[width=85mm,angle=0]{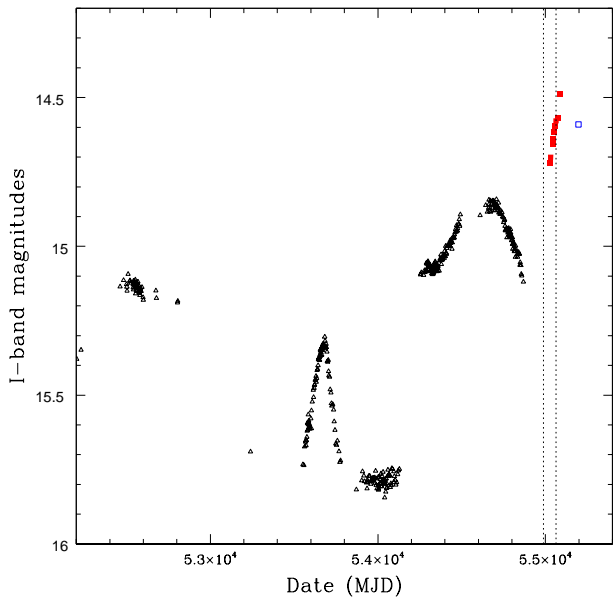}
  \caption{Combined OGLE III, Faulkes Telescope (FT) south and SAAO 1.9m I-band light curve of the optical counterpart in SXP11.5 (open black triangles, closed red squares and open blue square respectively). The FT I-band data have been transformed into the OGLE I-band for direct comparison of the magnitude of the star. The vertical dashed lines indicate the start and end of the Type II X-ray outburst as seen by \textit{INTEGRAL} and \textit{RXTE}; this time period spans approximately two orbital cycles, illustrating the scale of the optical variability.\label{fig:ogle}}
\end{figure}

\begin{figure}
 \includegraphics[scale=0.5]{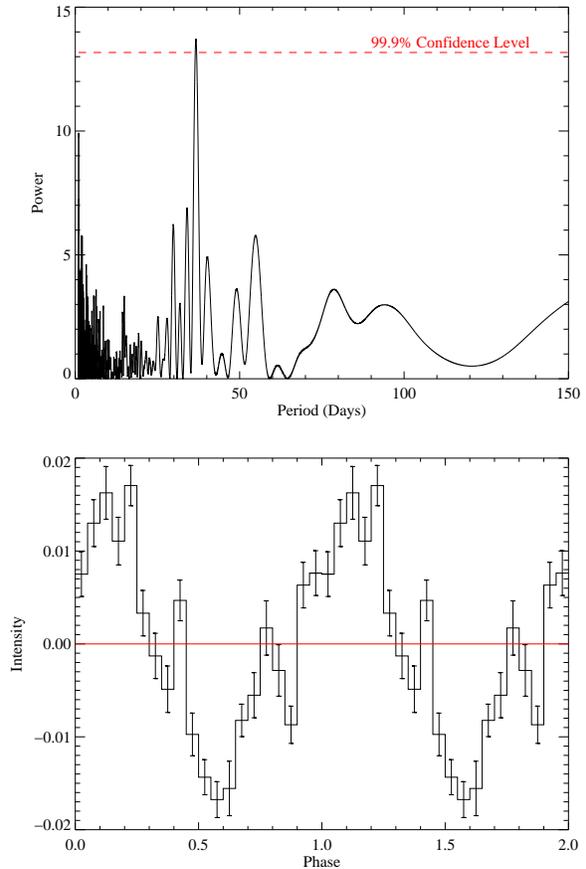}
  \caption{Lomb-Scargle periodogram of the detrended section of the OGLE III light curve between MJD 53400 and MJD 54200 (top). A peak at 36.70 $\pm$ 0.03d is seen above the 99.9\% significance level, verifying the orbital period predicted by the X-ray data. This period was only detected in this section of the light curve when the disk was seemingly at its lowest flux level. Analysis of other individual sections and the light curve as a whole did not reveal any significant periods. The phase-folded light curve (bottom) is folded such that a phase of 1.0 in the figure corresponds to periastron of the NS based on the result of our X-ray fit.\label{fig:optLS}}
\end{figure}

Fig. \ref{fig:ogle} shows the full 12yr optical light curve of SXP11.5 from OGLE monitoring (open black triangles) and coverage from Faulkes Telescope (FT; closed red squares) south and the 1.9m, SAAO telescope (open blue square). The FT I-band data were calibrated onto the OGLE I-band using 18 field stars from the OGLE database that show no variability over the set of measurements. The OGLE data are sparse compared to the normal coverage of the SMC because this system occasionally falls onto a gap in the detector chip which is dependent on its orientation angle. Despite this, the data presented were detrended by subtraction of a 50 day moving mean model to remove large amplitude variability, allowing timing analysis to be performed. A Lomb-Scargle periodogram of the detrended light curve shows the presence of a 36.70 $\pm$ 0.03d periodicity above the 99.9\% significance level, as shown in Fig. \ref{fig:optLS}. This periodicity was only detected in certain sections of the light curve, but confirms the orbital period predicted from the X-ray data. Fig. \ref{fig:optLS} also shows the phase-folded light curve folded at the detected period and phase alligned to the ephemeris given in Table \ref{tab:orbit}. The profile seems to be quite sinusoidal, being brightest just after periastron and faintest just after apastron, suggestive of a slight lag between the periodic brightening of the disk with the phase of the NS. The narrow peak and trough at phases of approximately 0.4 and 0.9 respectively are somewhat more peculiar; they may suggest something is happening half way between the optical minimum and maximum. However, on folding the light curve at various phases or with different bin sizes these features come and go, suggesting that they may not be a real physical feature of the system. There are no MACHO or OGLE II data of this source. The OGLE, FT and SAAO data were taken before, during and after the X-ray outburst respectively (see discussion section for interpretation). The orbital period found here was also used as a frozen parameter in a second orbital model fit to the X-ray data. This did improve slightly the errors on the parameters presented in Table \ref{tab:orbit} but only on the order of a few percent.

\begin{table}
  \caption{IR and optical data of the counterpart to SXP11.5.\label{ta:irsf}}
  \begin{tabular}{@{}ccccc@{}}
  \hline
   Catalogue & Date (MJD) & J & H & $K_{s}$\\
  \hline
   2MASS$^{1}$ & 51035 & $14.16\pm0.03$ & $13.70\pm0.04$ & $13.52\pm0.04$\\
   SIRIUS$^{2}$ & 52894 & $14.21\pm0.08$ & $13.75\pm0.06$ & $13.74\pm0.04$\\
  \hline
   Telescope & Date (MJD) & J & H & K$_{s}$\\
  \hline
   IRSF & 55196 & $14.23\pm0.04$ & $13.75\pm0.03$ & $13.61\pm0.04$\\
  \hline
   Telescope & Date (MJD) & B & V & I\\
  \hline
   1.9m & 55200 & $14.91\pm0.31$ & $14.91\pm0.11$ & $14.59\pm0.10$\\
  \end{tabular}\\
 \begin{flushleft}
   $^{1}$\citet{skr06}, $^{2}$\citet{kato07}.
\end{flushleft}
\end{table}

Photometric data taken on the IRSF and the SAAO 1.9m telescope are shown in Table \ref{ta:irsf}, along with measurements previously published in catalogues. The IRSF Magellanic Clouds Point Source catalogue by \citet{kato07} was used to calibrate the IRSF data. The Magellanic Clouds Photometric Survey (MCPS) catalogue by \citet{zar02} was used to calibrate the 1.9m optical data. The optical photometry is not as good as the IR as can be seen from the associated errors. This is due to the poor seeing at optical wavelengths during the observations, and variability in the conditions during the long integration times required because of a full moon. Also, the smaller field of view of the SAAO CCD compared with the SIRIUS CCD meant there were fewer stars which could be used to perform the calibration. A pipeline provided by SAAO was used to calculate the PSF magnitude of the optical images after performing flat fielding. The error is calculated by taking the average of the square of the magnitude errors from the MCPS catalogue, for the stars used for calibration, and the standard deviation of the stars from the fitted line. The final error is $\sqrt{\sigma_\mathrm{avg,catalogue}^{2} + \sigma_\mathrm{fit}^{2}}$. Data from the IRSF Magellanic Clouds Point Source catalogue and 2MASS catalogue \citep{skr06} are presented as reference to past IR observations of this source. Unfortunately, the IR data are not directly comparable to the I-band data in the OGLE light curve as both measurements fall into gaps in the data and as such we cannot say if the NIR bands follow what is happening in the I-band. However, we can make two cautious observations; firstly, that there is little variation in the J, H \& K band values presented in Table \ref{ta:irsf} and, secondly, that at the time of the SIRIUS catalogue measurement the I-band seems to be much fainter than at the time of the IRSF measurement (assuming the I-band at MJD 53000 is around 15.2 magnitudes in comparison to the present value of 14.6). This suggests that there may be a change in the (I-K) colour in this system. If true then this is indicative of a changing temperature within the optically thick circumstellar disk.

\subsection{Spectroscopy}

SXP11.5 was observed during the X-shooter Science Verification (SV1) phase in August 2009 (MJD 55055; $\sim$10d before the X-ray outburst ended), with a total integration time of 1200s\footnote{Program ID 60.A-9439(A)}. The spectra were reduced with a beta version of the ESO X-shooter pipeline \citep{gold06}, which uses subtraction of the sky lines based on the procedures developed by \citet{kel03}. The pipeline reduction used calibration spectra taken during the commissioning run for order location and tracing, flat fielding and wavelength calibration. The final product from the pipeline is an extracted 2D, wavelength calibrated, rectified spectrum with orders combined using a weighting scheme. For further data processing and analysis we used a combination of IDL programs and standard IRAF tools. The spectra were then smoothed with a boxcar average of 7, normalised to remove the continuum and then shifted by −150 km s$^{-1}$ \citep{all73} to account roughly for the recession velocity of the SMC and hence to place spectral features at approximately the correct wavelengths. These spectra are presented in Figs. \ref{fig:xshoot} and \ref{fig:halpha}. The UVB arm spectrum is used in the next section to spectrally classify SXP11.5, whilst the VIS arm spectrum is used to measure the H$\alpha$ equivalent width just before the outburst came to an end. This is discussed and compared to a SAAO H$\alpha$ measurement taken months later in section 6.

\section{Spectral Classification}

\begin{figure*}
 \includegraphics[width=130mm,angle=90]{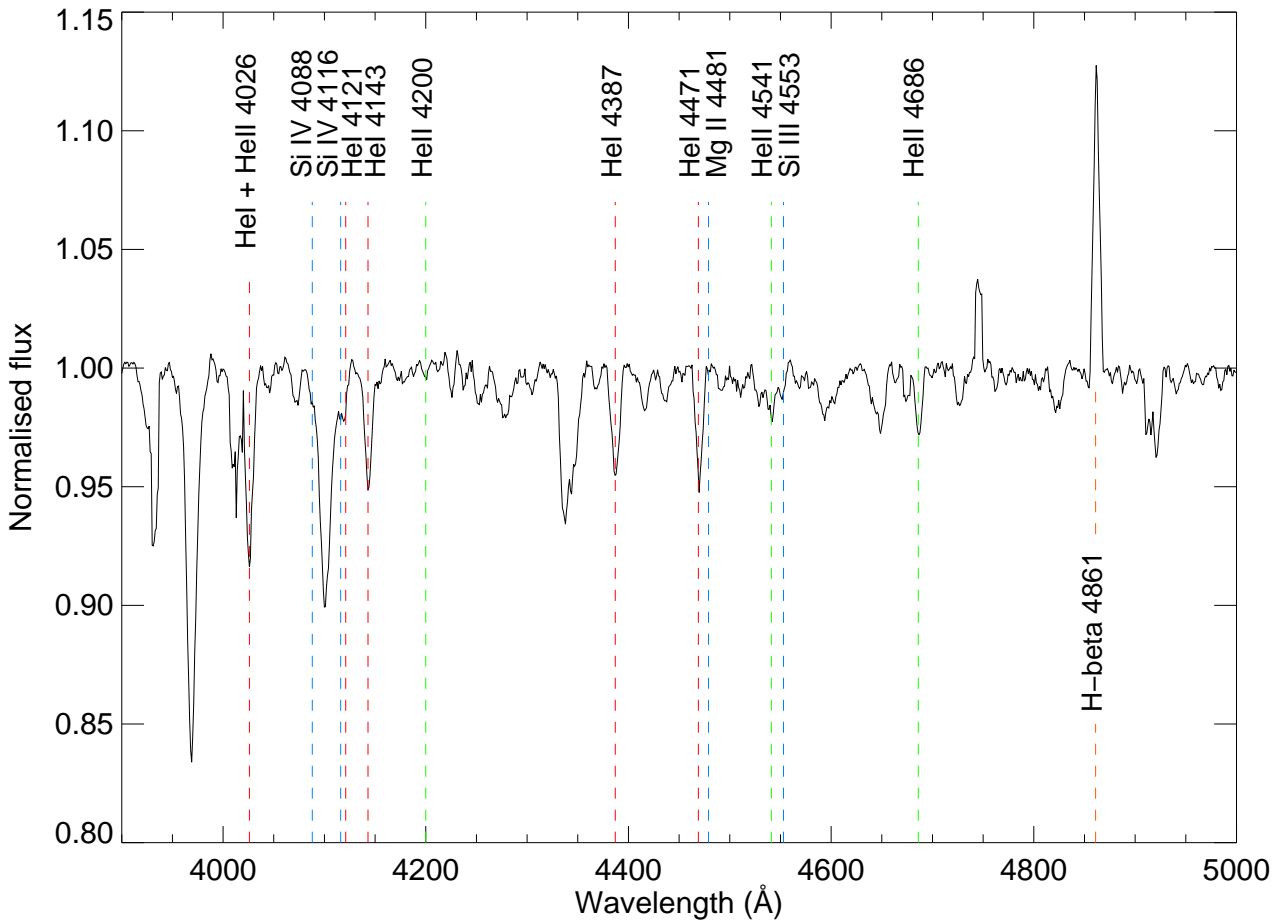}
  \caption{Spectrum of SXP11.5 taken with the UVB arm of the X-shooter spectrograph in the wavelength range 300-550nm. The spectrum has been normalised to remove the continuum and the data have been redshift corrected by -150 km s$^{-1}$ to account for the recession of the SMC. Overplotted are various atomic transitions that are significant in the spectral classification of an early type star at their rest wavelengths; He II, He I and metal transitions are in green, red and blue respectively.\label{fig:xshoot}}
\end{figure*}

Classification of Galactic Be stars relies on using the ratio of many metal-helium lines \citep{wal90}. However, this type of classification, based on the Morgan-Keenan (MK; \citealt{mkk43}) system, is particularly difficult in the low metallicity environment of the SMC because these metal lines are either very weak or not present at all. Using high signal-to-noise ratio spectra of SMC supergiants, \citet{len97} devised a system for the classification of stars in the SMC that overcomes the problems with low metalicity environments. This system is ‘normalized’ to the MK system such that stars in both systems exhibit the same trends in their line strengths. Another difficulty is that the Balmer lines in particular will be rotationally broadened due to the high rotational velocities of Be stars and hence may obscure any comparisons to closely neighbouring lines. We have thus used the classification method as laid out in \citet{len97} and utilised further in \citet{evans04}. For the luminosity classification we have adopted the classification method set out in \citet{wal90}.

The classification of SXP11.5 has been made using the X-shooter UVB arm spectrum presented in Fig. \ref{fig:xshoot}. Immediately obvious is the presence of ionised Helium in the spectrum: He II $\lambda\lambda$4686, 4541 are present, meaning the spectral type must be B0 or earlier (\citealt{len97}; \citealt{evans04}). He II $\lambda$4200 is weak, and much weaker than He I $\lambda$4143, meaning it is later in type than O9. There may be evidence of some Si lines in the spectrum which could stretch the classification to B1, although these are very difficult to confirm above the general SNR of the spectrum. Given the presence of He II $\lambda$4686 it would seem that, even if the metal lines are supressed by the low metallicity of the SMC or masked by rotational broadening of the Balmer and helium lines, a spectral type as late as B1 is unlikely. \citet{wal90} also present spectra with apparent Si lines and classify them as B0, indicating that a classification of B0 would be correct in this case if the Si lines are present. Given that the metal lines in the spectrum are not so obvious, we were limited to using the He I $\lambda$4121/He I $\lambda$4143 ratio to determine the luminosity class \citep{wal90}. This line ratio strengthens towards more luminous stars, suggesting that this star has a luminosity class of IV-V. We have also performed a check on this luminosity classification by comparing the absolute magnitude of the source in the V-band with a distance modulus for the SMC of 18.9 \citep{hhh03}, to determine whether the absolute magnitude of the source is consistent with the estimated luminosity class for this spectral type (using absolute magnitudes for OeBe stars from \citealt{weg06}). A V-band magnitude of 14.9 (Table \ref{ta:irsf}) confirms a luminosity class of IV-V for a spectral type of O9.5-B0, as a star of higher luminosity class would need to be of a later spectral type. Our spectral classification agrees with recent work done by \citet{masetti10} on the SAAO 1.9m telescope, although those authors give a luminosity class of III which may be too early based on the available line ratios in the spectrum.

\section{Discussion}

\subsection{X-ray behaviour}

Strong triple peaked structure is visible in the X-ray pulse profile when the 3rd harmonic was strongest (top panel Fig. \ref{fig:folded_lc}). This triple structure remained constant for several observations whilst the outburst was at its most intense. Only once the flux had significantly dropped did the pulse shape begin to evolve into a more familiar single peaked structure (bottom panel Fig. \ref{fig:folded_lc}). At this point the 3rd harmonic fades and the fundamental begins to dominate the pulsed emission. The energy dependent pulse profiles plotted in Figs. \ref{fig:rxte:pulseprofile} and \ref{fig:xmm:pulseprofile} demonstrate a possible change in the X-ray emission at higher energies. The \textit{XMM-Newton} profile seems to show the shape of the folded light curves are very similar within the 0.2--10 keV energy range (although we are aware that some information is being lost here due to the none-detection of the 3rd harmonic). This was also the case for the \textit{RXTE} observations in which the 3--10 keV light curve was split into smaller energy bands (these have not been presented here). However, moving to higher energies in the \textit{RXTE} profiles we see a change occuring around 10 keV. Below this energy, the pulse profile suggests a combination of a pencil beam from one pole and a fan beam from the other pole, with an orientation that places two sides of the fan beam and the pencil beam approximately one third of a phase apart. The smaller of the three peaks is likely the pencil beam in this picture, being a phase of 0.5 apart from the centre of the fan beam 'double' peak. The hard energy profile shows evidence for a number of possibilities; the pencil beam has increased in strength, becoming comparible to the fan beam. The pencil beam may be narrower, although it remains 0.5 phase away from the centre of the other 2 peaks. The fan beam may have a narrower opening angle at higher energies given the apparant infilling of the gap between the two peaks. At the moment, these are all speculations based on observation and thus more detailed modelling is required to take this discussion further.

The orbital parameters in Table \ref{tab:orbit} are typical of those seen in other BeXRB systems \citep{oka01}. Most Be systems have P$_{orb} \sim$ 10-100\,d and orbital eccentricities of between 0.3-0.5 \citep{bild97}. These parameters also place SXP11.5 firmly in the BeXRB regime of the Corbet diagram \citep{cor86}, lending support to the optical determination that this source is a BeXRB and not a supergiant system.

\begin{figure}
 \includegraphics[width=85mm,angle=0]{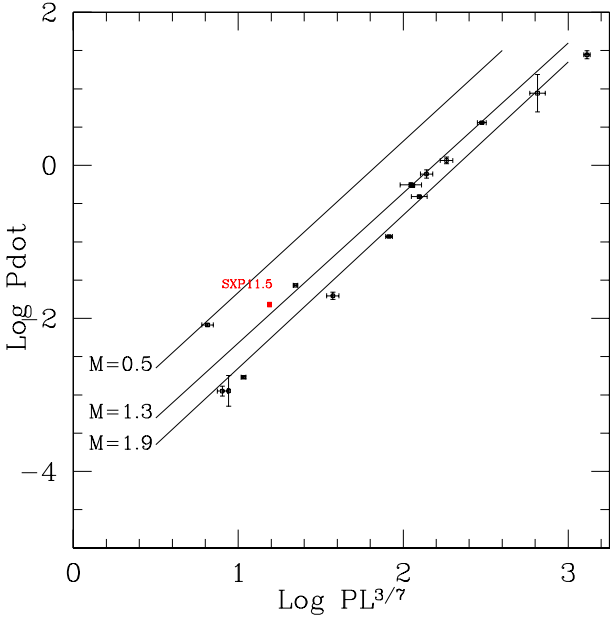}
  \caption{The distribution of SMC BeXRB pulsars on the $\dot{P}$ vs $PL^{\frac{3}{7}}$ plane adapted from \citet{coe10b}, where $\dot{P}$ is the spin-up in sec/yr, $P$ is the spin period in seconds and $L$ is the X-ray luminosity in units of $10^{37}$ ergs/s. SXP11.5 is shown to sit in the low period end of the distribution, confirming that the luminosity, pulse period and spin-up calculated in this analysis agree with the equations orginally presented in \citet{ghosh79} and that this source is, in this sense, normal with respect to the general distribution of SMC pulsars.\label{fig:lamb}}
\end{figure}

\citet{ghosh79} show that there is a relationship for binary pulsars such that -$\dot{P}$ $\mathrm\alpha$ $PL^{\frac{3}{7}}$ for a given NS mass and magnetic moment, where $\dot{P}$ is the spin-up in sec/yr, $P$ is the spin period in seconds and $L$ is the X-ray luminosity in units of $10^{37}$ ergs/s. This relationship has since been verified by \citet{coe10b} using a much larger database of SMC binary pulsars. The spin period and $\dot{P}$ from our fit were used along with the highest \textit{RXTE} luminosity to place this new system onto the plot of SMC pulsars presented in \citet{coe10b}. The result is shown in Fig. \ref{fig:lamb}. The position of this source on the plot seems to follow the general distribution of SMC pulsars and thus agrees with the theoretical laws of binary accretion.

\subsection{Optical behaviour}

Due to the lower temperatures believed to be present in the circumstellar environment, the photmetric I-band data presented in Fig. \ref{fig:ogle} are a direct measure of the circumstellar emission. These data show that this system has a highly variable circumstellar disk and that the X-ray outburst was triggered by the disk being in an exceptionally large state, relative to its base level. The vertical dashed lines in Fig. \ref{fig:ogle} show the start and end points of the X-ray outburst in relation to the optical activity. A single observation at SAAO in January 2010 (MJD 55200) shows that the disk has started to decrease in size, although the X-ray outburst had finished before the disk had returned to the level at which the X-ray emission began. This could be evidence of a lag between the X-ray and optical emission in this system. Another puzzle that these data introduce is why we only see the orbital modulation in the light curve at the very lowest flux level. This would suggest that the variability is coming from the stellar atmosphere as the disk would seem to be at a minimum. In contrast \citet{schurch10} only detect the orbital modulation from SXP18.3 at the very brightest part of the detrended optical light curve, suggesting that in this system, the variability is coming from the circumstellar disk and not the stellar atmosphere.

\begin{figure}
 \includegraphics[width=67mm,angle=90]{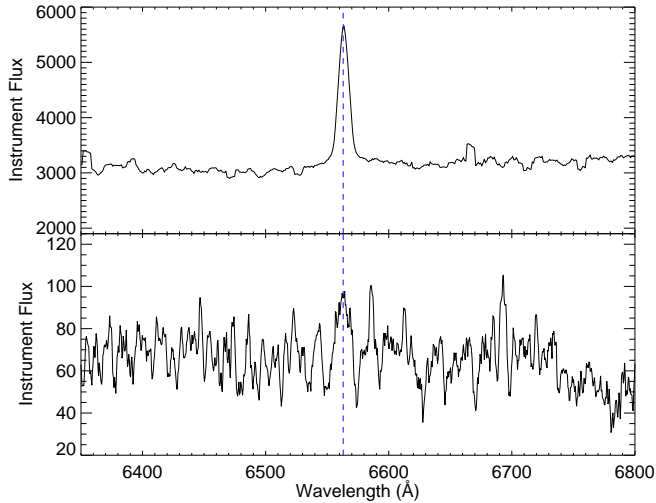}
  \caption{Spectra of SXP11.5 taken in the wavelength range of H$\alpha$ at the VLT (top) and SAAO,1.9m (bottom). The VLT spectrum was taken on 2009-08-12 (MJD 55055), just before SXP11.5 fell below the detection threshold of \textit{RXTE}, therefore showing the disk during an X-ray active phase. The SAAO spectrum was taken on 2009-12-12 (MJD 55177), four months after the X-ray outburst ended.\label{fig:halpha}}
\end{figure}

It is currently believed that the amount of red and IR emission from the circumstellar disk is directly related to the strength of emission lines in this regime. Fig. \ref{fig:halpha} shows two spectra of the H$\alpha$ region taken during and many months after the X-ray outburst. The top panel is a VLT/X-shooter VIS arm spectrum showing clearly H$\alpha$ being in emission. This corresponds to the peak of the I-band light curve shown in Fig. \ref{fig:ogle}. Equivalent width measurements of the line profile give a value of -8.0$\pm$0.5\.{A}. The bottom panel is a spectrum taken on the SAAO, 1.9m telescope 4 months after the outburst had ended. This observation was near-simultaneous to the SAAO I-band measurement plotted in Fig. \ref{fig:ogle}. The poor SNR in this spectrum made it difficult to make a precise equivalent width measurement, the best estimate being -3 $\pm$ 2, so we cannot say by how much the disk has reduced in size. However, we can conclude that both the H$\alpha$ and I-band measurements suggest the disk is shrinking. Comparision to H$\alpha$ measurements of other BeXRBs shows that, even during a long Type-II X-ray outburst, the disk in this system was comparably quite small (we have equivalent width measurements of more than -55\.{A} in some other systems; \citealt{coe05}), although it seems not abnormally small based on the relationship presented in \citet{ant09}; those authors present the H$\alpha$ equivalent width of a sample of BeXRBs against their orbital period and show a linear trend. Although SXP11.5 lies below the best fit line of this trend, it is not a complete outlier and may just lie within the distribution. Unfortunately a more quantitive handle on the IR excess seen in the photometry is needed to help verify the true size of the disk relative to other systems in the SMC. So given only a small circumstellar disk to accrete from, how does accretion continue for over two months in this system? The most plausible explanation seems to be that an accretion disk has formed around the NS, allowing accretion to continue after the NS has moved away from the circumstellar disk which may not, in this case, be filling, or be near to filling the orbit of the NS. However, it is not easy to detect the presence of an accretion disk in HMXBs and so an unusual orbit or other such geometries cannot be ruled out as an explanation of what is seen, although this seems unlikely based on the fairly standard parameters obtained in the orbital fit of the X-ray data. It is also worth noting that the narrow, single peaked H$\alpha$ and H$\beta$ emission lines seen in the spectra presented here suggest the disk has a low inclination angle to the observer \citep{str30}. Comparing this to the value of a$_{x}$sin\textit{i} presented in Table \ref{tab:orbit} suggests that the semimajor axis of the disk could be over 300 light-s (for \textit{i} $<$ 30$^{\circ}$), approximately 18 R$_{\star}$ assuming a B0 star of radius equal to 7 R$_{\odot}$.

\begin{figure}
 \includegraphics[width=90mm,angle=0]{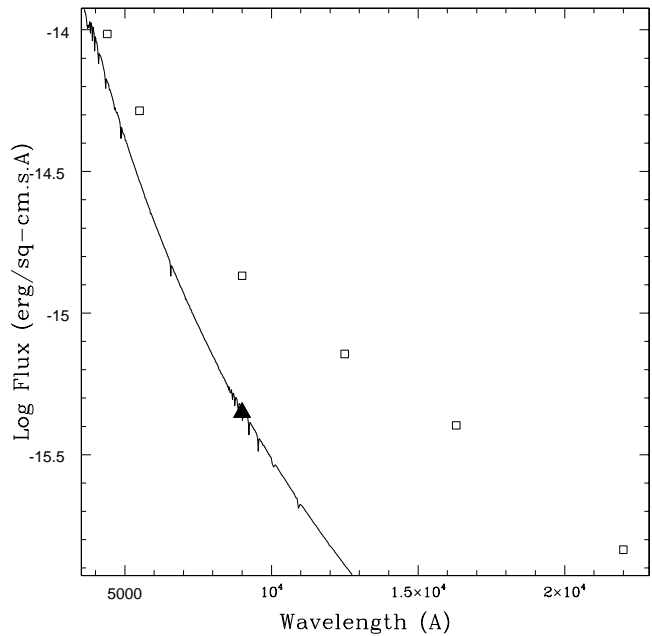}
  \caption{The optical \& IR photometry (open squares) from MJD 55196-55200 dereddened and compared to a Kuruscz model atmosphere for a B0V star (continuous line). The filled triangle indicates the lowest OGLE III I-band point which has been used to normalise the Kuruscz model - see text for details. This demonstrates the extent of the continuum emission from the circumstellar disk in this system.\label{fig:excess}}
\end{figure}

In order to explore the extent of the emission from the circumstellar disk, the contemporaneous optical \& IR photometric data from MJD 55196-55200 shown in Table \ref{ta:irsf} were used to compare with the predicted stellar emission. The photometric data were first dereddened by the established value to the SMC of E(B-V)=0.08 \citep{si91}. These were then compared to a Kurucz model atmosphere \citep{kur79} for a B0V star ($T_{eff}$=30,000K and log g=4.0). To make the comparison, the model atmosphere was normalised to the lowest I-band point in the OGLE III data (taken around MJD 54000) on the assumption that this represents the occasion on which any contribution from the circumstellar disk is at its lowest. The results are shown in Fig. \ref{fig:excess} from which it is immediately apparent that at the time of the optical \& IR photometry there was a clear disk contribution across the whole of the optical \& IR regime. Even the B \& V points clearly contain some element of such a component. In trying to quantify this excess, we normalised the Kuruscz model to the distance estimate of the SMC (60 kpc) and found that it lay very close to the I-band normalised model. Working from this we converted the I-band flux at this base level into a distance estimate, assuming that all the flux is coming from the star. We find a distance to SXP11.5 of 71$\pm$5 kpc. The error comes from calculating the maximum and minimum distance based on a range of E(B-V) values. Of course, this value is very dependant on obscuration localised to the SMC itself and should be treated with some caution. What this calculation shows is that it is very likely that the I-band measurement used in the normalisation represents very closely the base level of the disk emission in this system. A H$\alpha$ spectrum is needed to confirm this prediction should the I-band magnitude ever return to this level.

\section{Conclusions}

We have presented the complete orbital solution to a newly discovered high-mass binary system in the SMC, making it one of the few well described systems known outside the Galaxy. These parameters are similar to other binary systems in the Galaxy and Magellanic Clouds and place SXP11.5 firmly in the BeXRB region of the Corbet diagram. Under huge accretion torques, it is shown that the NS is being spun up and that this satisfies current theoretical predictions of the relationship between spin-up, spin period and luminosity in binary systems. However, it is clear that there is still much to learn about the accretion process itself as shown by the complex triple peaked structure and evolution of the pulse profiles, which are still poorly understood.

Spectral analysis of the companion has allowed a classification of O9.5-B0 IV-V to be made. Timing analysis of the optical light curve confirms the orbital period detected in the X-ray. Contemporaneous H$\alpha$ spectra show, along with the optical light curve, that the circumstellar disk is shrinking in the time since the X-ray outburst. Although there is evidence that the disk is unusually small compared to other BeXRBs, making the extended period of accretion seen in this outburst difficult to explain. We show that the lowest flux seen in the optical light curve is likely to be from a period in which the Be star has entirely lost its disk and make a distance estimate to the source based on this assumption. H$\alpha$ spectroscopy will be necessary to confirm this prediction should the I-band flux ever return to this level.

SXP11.5 has proven to be a very interesting addition to the SMC binary population; on one hand, it seems to be a very normal system showing orbital parameters and a spectral type that match very well with previously studied systems. However, it also shows peculiarities, such as showing a small $H\alpha$ equivalent width relative to other systems with similar orbital periods. Should this system go into another X-ray outburst in the future, further simultaneous optical and X-ray measurements are essential to uncovering where this system lies in the overall population of SMC BeXRB systems.

\section*{Acknowledgements}

LJT is supported by a Mayflower scholarship from the University of Southampton. ABH is funded by contract ERC-StG-200911 from the European Community. DP acknowledges support from a Dorothy Hodgkin Postgraduate Award. BvS is funded by the South African Square Kilometre Array Project. This paper uses observations made at the South African Astronomical Observatory (SAAO). The authors are grateful for the assistance of D.J. Wium with the photometric observations performed on the SAAO 1.9m telescope. The OGLE project is partially supported by the Polish MNiSW grant N20303032/4275. LJT would like to thank the X-shooter team at ESO for providing a good beta-version pipeline and several helpful suggestions during the reduction of the SV1 data presented in this paper. We would like to thank the anonymous referee for their positive and swift feedback and constructive comments.

\bsp

\label{lastpage}

\end{document}